\documentclass[hyper]{JHEP3}
\usepackage{graphicx}
\usepackage{cite}
\def\beq{\begin{equation}}
\def\eeq{\end{equation}}
\def\beqn{\begin{eqnarray}}
\def\eeqn{\end{eqnarray}}
\def\nn{\nonumber\\}

\title{Reconstructing particle masses from pairs of decay chains}

\author{Mihoko M.\ Nojiri$^{123}$, Kazuki Sakurai$^{45}$ and Bryan R.\ Webber$^4$\\
$^1$ Institute for the Physics and Mathematics of the Universe, The University of Tokyo, 
Chiba 277-8568, Japan\\
${}^2$Theory Group, KEK, 1-1 Oho, Tsukuba, Ibaraki 305-0801, Japan\\
${}^3$The Graduate University for Advanced Studies (SOKENDAI),1-1 Oho,
Tsukuba, Ibaraki 305-0801, Japan\\
$^4$Cavendish Laboratory, J.J.\ Thomson Avenue, Cambridge CB3 0HE, UK\\
$^5$Department of Applied Mathematics and Theoretical Physics\\ 
Wilberforce Road, Cambridge CB3 0WA, UK\\
\\
       E-mail: \email {nojiri@post.kek.jp,sakurai@hep.phy.cam.ac.uk,webber@hep.phy.cam.ac.uk}
        }

\received{\today}               
\accepted{\today}               

\preprint{Cavendish-HEP-10/09\\DAMTP-2010-38\\IPMU 10-0081 \\KEK-TH-1362}

\abstract{
A method is proposed for determining the masses of the new particles
$N,X,Y,Z$ in collider events containing a pair of effectively
identical decay chains $Z\to Y+$ jet, $Y\to X+l_1$, $X\to N+l_2$,
where $l_1,l_2$ are opposite-sign same-flavour charged leptons and $N$
is invisible.  By first determining the upper edge of the dilepton
invariant mass spectrum, we reduce the problem to a curve for
each event in the 3-dimensional space of mass-squared differences.
The region through which most curves pass then determines the unknown
masses.  A statistical approach is applied to take account of
mismeasurement of jet and missing momenta. 
The method is easily visualized and rather robust against
combinatorial ambiguities and finite detector resolution.
It can be successful even for small event samples,
since it makes full use of the kinematical information from every event.  
}


\begin{document} 

\section{Introduction}
\label{sec:intro}
One of the principal objectives of the ongoing experiments at the
Large Hadron Collider is the discovery of new physics beyond the
Standard Model.   Many models of BSM physics predict a rich spectrum
of new particles with sequential decays into chains of other
new particles plus visible jets and leptons.  Typically the endpoint
of the chain is a new stable invisible particle that is a dark matter
candidate.  Important examples are the squark decay chain in
supersymmetric models,
\beq\label{eq:chain_susy}
\tilde q\to \tilde\chi^0_2 + q\,,\;\tilde\chi^0_2 \to \tilde\ell^\pm+\ell^\mp\,,\;
\tilde\ell^\pm\to\tilde\chi^0_1+\ell^\pm\;,
\eeq
where the neutralino $\tilde\chi^0_1$ is the lightest supersymmetric particle (LSP),
and the excited quark decay in models with universal extra dimensions,
\beq\label{eq:chain_ued}
q^*\to Z^* + q\,,\; Z^* \to \ell^{*\pm}+\ell^\mp\,,\;
\ell^{*\pm}\to\gamma^*+\ell^\pm\;,
\eeq
where the photon excitation $\gamma^*$ is the lightest Kaluza-Klein particle.

If such decay chains do indeed occur at the LHC, the most urgent and challenging
task will be to determine the masses and other properties of
the new particles involved.  Many approaches to the mass determination problem have been
proposed,\footnote{For a recent review, see~\cite{Barr:2010zj}.}
based mainly on the measurement of endpoints, kinks or other features in the
distributions of invariant masses or specially constructed
observables, or on explicit solution for the unknown masses using multiple events.

The present paper investigates a combination of the explicit-solution
and endpoint methods for processes in which there are two effectively
identical three-step decay chains, such as first- and second-generation squark pair
production and decay as in (\ref{eq:chain_susy}).\footnote{The
  explicit-solution method using pairs of events has been applied to the
  same class of processes in refs.~\cite{Cheng:2008mg,Cheng:2009fw}.}
The method is a development of the approaches in
refs.~\cite{Kawagoe:2004rz,Nojiri:2007pq,Webber:2009vm},
combining kinematic fitting with endpoint information to
represent the possible mass solutions for each individual
event as a curve in a three-dimensional space of mass-squared
differences.   For exact kinematics, the curves of different events
all intersect at the unique correct solution point.  In the presence
of combinatorial ambiguities, measurement errors and mass variations,
the region where the density of curves is highest gives the best estimate of the
masses.  Unlike pure endpoint or kink methods, this approach makes full
use of the kinematical information from every event, no matter where
it may lie in phase space.  Determination of the four sparticle masses
and reconstruction of the LSP momenta then appears possible even with
small event samples.

In Section 2 we present the general method and then in Section 3 we show
results for a number of SUSY model points that lead to the decay chain
(\ref{eq:chain_susy}). Our conclusions are summarized in Section 4.
  
\section{Method}
\label{sec:method}
\begin{figure}\begin{center}
\includegraphics[width=70mm]{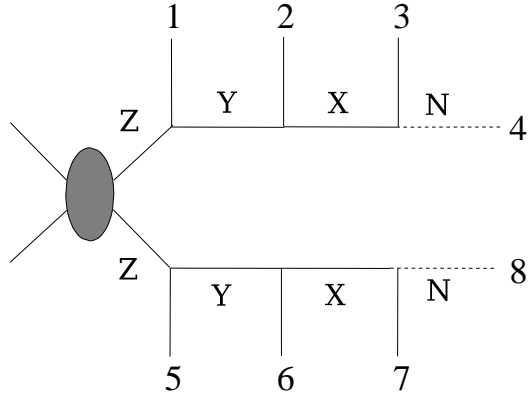}
\caption{Event topology}\label{fig:event}
\end{center}\end{figure}
Consider the double decay chain in fig.~\ref{fig:event}.
The 4-momenta in the upper chain should satisfy
\beqn
(p_1+p_2+p_3+p_4)^2 &=& m_Z^2\nn
(p_2+p_3+p_4)^2 &=& m_Y^2\nn
(p_3+p_4)^2 &=& m_X^2\nn
p_4^2 &=& m_N^2\;.
\eeqn
Defining the mass-squared differences
\beqn\label{eq:Mdef}
M_1 &=& m_Z^2-m_Y^2>0\nn 
M_2 &=& m_Y^2-m_X^2>0\nn 
M_3 &=& m_X^2-m_N^2>0\;,
\eeqn
the first three equations give linear constraints on the invisible 4-momentum $p_4$:
\beqn\label{eq:upconstr}
-2p_1\cdot p_4 &=& 2p_1\cdot p_2 +2p_1\cdot p_3+m_1^2-M_1\equiv S_1\nn
-2p_2\cdot p_4 &=& 2p_2\cdot p_3 +m_2^2-M_2\equiv S_2\nn
-2p_3\cdot p_4 &=& m_3^2-M_3\equiv S_3\;.
\eeqn
Similarly for the lower chain
\beqn
-2p_5\cdot p_8 &=& 2p_5\cdot p_6 +2p_5\cdot p_7+m_5^2-M_1\equiv S_5\nn
-2p_6\cdot p_8 &=& 2p_6\cdot p_7 +m_6^2-M_2\equiv S_6\nn
-2p_7\cdot p_8 &=& m_7^2-M_3\equiv S_7\;.
\eeqn
We also have the missing transverse momentum constraints
\beqn
p_4^x+p_8^x &=& p^x_{\rm miss}\equiv S_4\nn
p_4^y+p_8^y &=& p^y_{\rm miss}\equiv S_8\;.
\eeqn
If we make an 8-vector of the invisible 4-momenta,
\beq
{\bf P} = (p_4^x,p_4^y,p_4^z,E_4,p_8^x,p_8^y,p_8^z,E_8)\;,
\eeq
then we have
\beq
{\bf A P} ={\bf S}
\eeq
where ${\bf A}$ is the 8$\times$8 matrix
\beq\label{eq:Adef}
{\bf A} = 2\left(\begin{array}{cccccccc}
p_1^x & p_1^y & p_1^z & -E_1 & 0 & 0 & 0 & 0 \\
p_2^x & p_2^y & p_2^z & -E_2 & 0 & 0 & 0 & 0 \\
p_3^x & p_3^y & p_3^z & -E_3 & 0 & 0 & 0 & 0 \\
 1/2 & 0 & 0 & 0 & 1/2 & 0 & 0 & 0\\
 0 & 0 & 0 & 0 & p_5^x & p_5^y & p_5^z & -E_5\\
 0 & 0 & 0 & 0 & p_6^x & p_6^y & p_6^z & -E_6\\
 0 & 0 & 0 & 0 & p_7^x & p_7^y & p_7^z & -E_7\\
 0 & 1/2 & 0 & 0 & 0 & 1/2 & 0 & 0
\end{array}\right)\;.
\eeq
Furthermore ${\bf S}$ may be written as
\beq
{\bf S} = {\bf B M} + {\bf C}
\eeq
where ${\bf M}=(M_1,M_2,M_3)$ is the 3-vector of mass-squared differences to be determined,
\beq
{\bf B} = \left(\begin{array}{ccc}
-1 & 0 & 0\\
0 & -1 & 0\\
0 & 0 & -1\\
0 & 0 & 0 \\
-1 & 0 & 0\\
0 & -1 & 0\\
0 & 0 &-1 \\
0 & 0 & 0
\end{array}\right)
\eeq
and
\beqn\label{eq:Cdef}
{\bf C} &=&
(2p_1\cdot p_2 +2p_1\cdot p_3+m_1^2, 2p_2\cdot p_3 +m_2^2, m_3^2,p^x_{\rm miss},\nn
&&\;\,2p_5\cdot p_6 +2p_5\cdot p_7+m_5^2,2p_6\cdot p_7 +m_6^2,m_7^2,p^y_{\rm miss})\;.
\eeqn
Hence the solution for the invisible 4-momenta is
\beq
{\bf P} ={\bf A}^{-1}{\bf S} = {\bf D M} + {\bf E}
\eeq
where ${\bf D} = {\bf A}^{-1}{\bf B}$ and  ${\bf E} ={\bf
  A}^{-1}{\bf C}$.

The  invisible 4-momenta also satisfy the quadratic constraints
\beqn\label{eq:MNs}
p_4^2 &=& P_4^2-P_1^2-P_2^2-P_3^2 =m_N^2\nn
p_8^2 &=& P_8^2-P_5^2-P_6^2-P_7^2 =m_N^2
\eeqn
where $m_N$ is an extra unknown, independent of ${\bf M}$. However, in
the case that the terminal pairs of visible decay products (2,3) and (6,7)
are opposite-sign same-flavour dileptons,  as in (\ref{eq:chain_susy})
or (\ref{eq:chain_ued}), we can reasonably expect
that the upper edge of the dilepton invariant mass spectrum will be
measured with good accuracy.  This quantity is given by
\beq
(m_{ll}^{\rm max})^2 \equiv M_L = (m_Y^2-m_X^2)(m_X^2-m_N^2)/m_X^2
=M_2M_3/(M_3+m_N^2)
\eeq
and hence
\beq\label{eq:mN2}
m_N^2=M_3(M_2/M_L-1)\;.
\eeq
Substituting this into eqs.~(\ref{eq:MNs}), we obtain a pair of
trivariate quadratic equations in $M_1,M_2,M_3$, whose real
solutions lie on a curve in the 3-dimensional space of those
variables.  The corresponding solutions for the new particle
masses are then given by eq.~(\ref{eq:mN2}) and
\beq\label{eq:mXYZ}
m_X^2 = m_N^2+M_3\,,\;\;\; 
m_Y^2 = m_X^2+M_2\,,\;\;\; 
m_Z^2 = m_Y^2+M_1\;.
\eeq
The limits $m_N^2>0$,  $m_Z^2< M_U$ (where $M_U= s/4$, $s$ being the
collider c.m.\ energy squared, or smaller, depending on the relevant
parton luminosities) imply that the solutions must lie within the region
\beqn\label{eq:Mregion}
0 &\leq M_3 &\leq M_U-M_L\nn 
M_L&\leq M_2 &\leq M_U/(1+M_3/M_L)\nn 
0  &\leq M_1 &\leq M_U-M_2(1+M_3/M_L)\;.
\eeqn
This is a finite region with volume
\beq
\frac 14 M_L\left[2 M_U^2\ln(M_U/M_L)-(3M_U-M_L)(M_U-M_L)\right]\;,
\eeq
which vanishes as $M_L\to 0$ or $M_L\to M_U$.

\section{Results}

As an illustration of the method, we apply it here to the process of squark-pair
production at the LHC ($pp$ collisions at 14 TeV centre-of-mass energy).  The SUSY mass
spectrum and decay branching ratios are taken to be those of CMSSM point
SPS 1a~\cite{Allanach:2002nj}.  The corresponding masses in the decay
chain  (\ref{eq:chain_susy}) are given in Table~\ref{tab:SPS1a}.
Events are generated using {\small HERWIG} version
6.5~\cite{Corcella:2000bw,Corcella:2002jc,Moretti:2002eu}. Some of the
squarks are produced directly and some come from gluino decay; the
production mechanism affects their momentum and rapidity distributions
but is otherwise irrelevant for our purposes. 

\begin{table}[!t]
  \centering
  \begin{tabular}{|c|c|c|c|}
    \hline $N$ & $X$ & $Y$ & $Z$ \\ 
    \hline $\tilde{\chi}_1^0$ & $\tilde{e}_R$ & $\tilde{\chi}_2^0$ & $\tilde{u}_L$ \\ 
    \hline 96 & 143 & 177 & 537 \\ \hline
  \end{tabular}
  \caption{Mass spectrum in GeV for Snowmass point SPS 1a}
  \label{tab:SPS1a}
\end{table}

Third-generation squarks are excluded, as their different masses prevent a good
fit with a single squark mass.  Experimentally, this would involve vetoing events
with a tagged $b$-jet.  At SUSY point SPS 1a only left-squarks have significant
branching ratios into the mode (\ref{eq:chain_susy}) and so the
left-right squark mass splitting is not a problem here.  The
$\tilde{d}_L-\tilde{u}_L$ mass difference is 5.8 GeV.  Therefore the
assumption that the masses in the two decay chains are identical
should be a good approximation.

To obtain the solution curve for each event, we proceed as follows.
As explained earlier, substitution of eq.~(\ref{eq:mN2}) into
eqs.~(\ref{eq:MNs}) gives a pair of trivariate quadratic equations in $M_1, M_2, M_3$.
Eliminating, for example, $M_3$ gives a quartic equation for $M_2$ as
a function of $M_1$.  For each real solution, $M_3$ is given
uniquely in terms $M_1$ and $M_2$.  Thus for each $M_1$ we obtain a
set of 0, 2 or 4 real solution points.  We divide the space of $(M_1,
M_2, M_3)$ into cells.  Scanning over $M_1$ gives a set of solution
points which occupy `hit' cells.  To find all hit cells we also scan
over $M_2$ and $M_3$, using permutations of the above procedure.  Hits
on already occupied cells are discarded.  The resulting set of points
provides an approximately uniform coverage of the solution curve. In
general, the solution curve may consist of one or two closed loops or
open segments with endpoints on the surface of the allowed region
(\ref{eq:Mregion}).

\begin{figure}\begin{center}
\includegraphics[width=50mm]{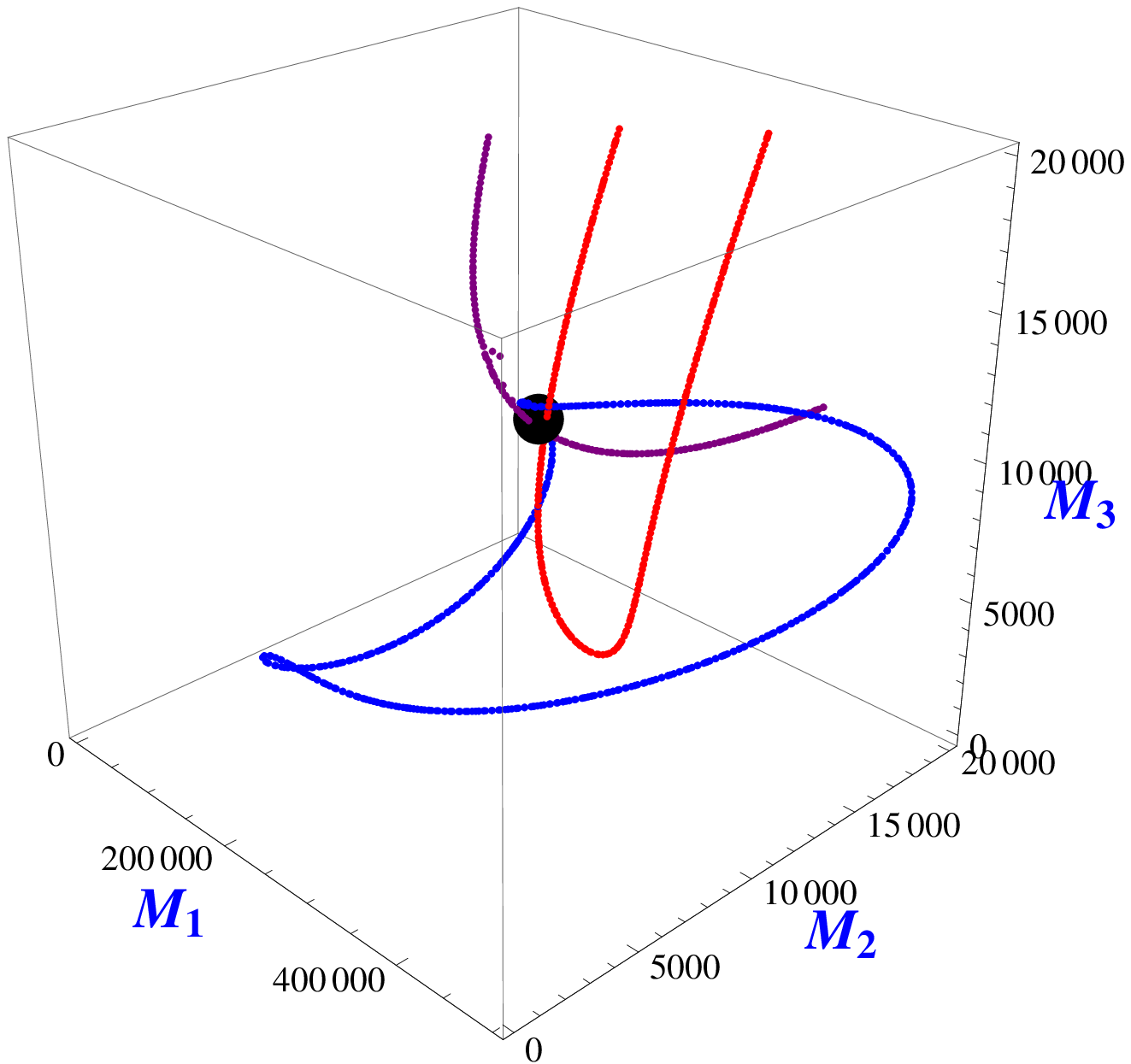}
\includegraphics[width=50mm]{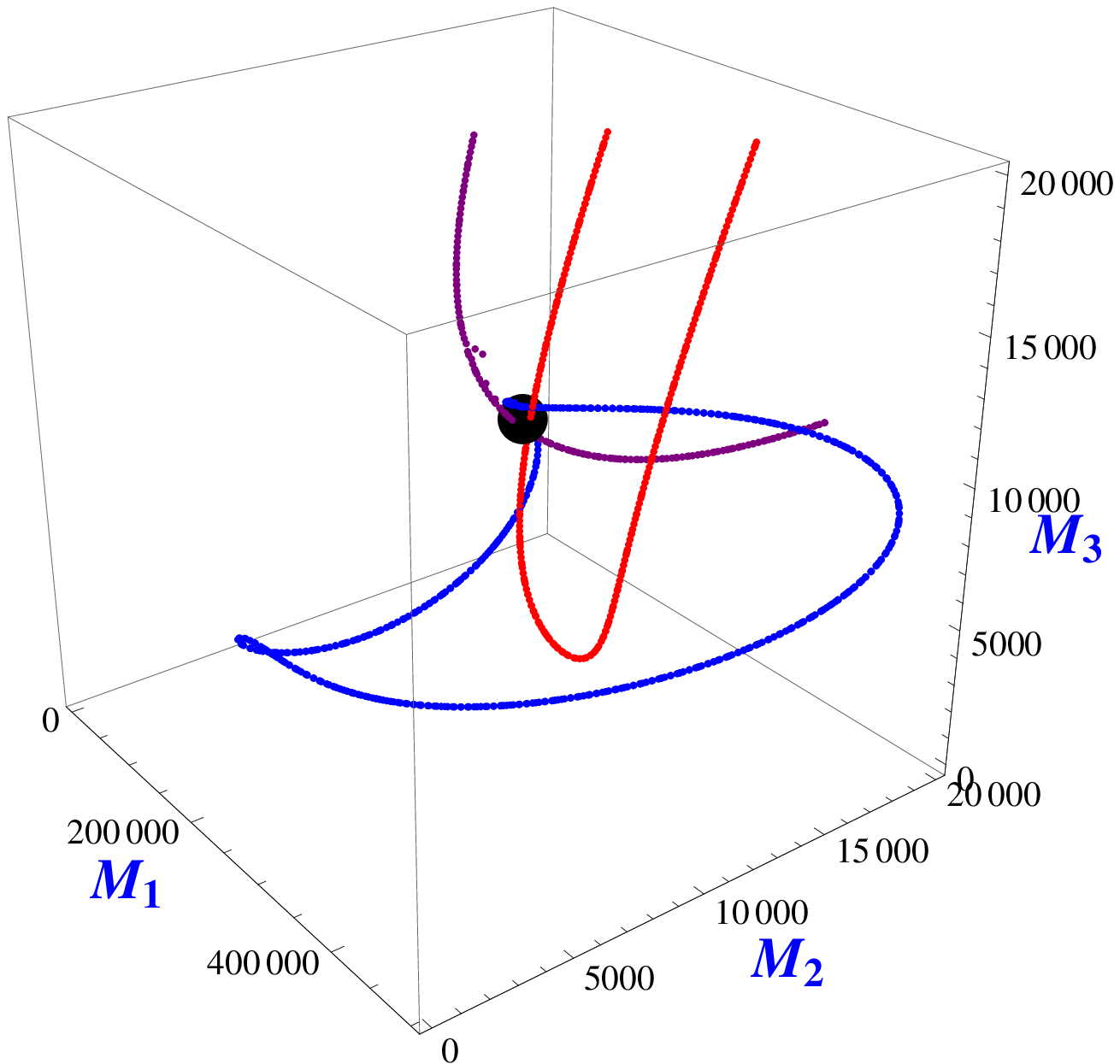}
\caption{Stereoscopic views of the true parton-level solution curves
  for three events. The ball shows the true mass
  point.}\label{fig:mass_tru3D}
\end{center}\end{figure}

Figure~\ref{fig:mass_tru3D} shows the parton-level
solution curves for three typical SPS1a events, using the correct
combinations of quarks and leptons in the decay chains.\footnote{The two
  images can be merged into a three-dimensional display by directing
  each eye at the corresponding image.}  The curves all pass
close to the ``true'' mass point (TMP)
\beq\label{eq:TMP}
M_1=257040\,,\;\;\; M_2=10880\,,\;\;\; M_3=11233\,,
\eeq
all in GeV$^2$, corresponding to the SUSY mass spectrum in
Table~\ref{tab:SPS1a}.  The curves do not precisely intersect, even
with exact kinematics, owing to Breit-Wigner smearing of unstable
particle masses.  However, we see that the density of solution curves
is high only in the vicinity of the TMP (\ref{eq:TMP}).

\begin{figure}\begin{center}
\includegraphics[width=50mm]{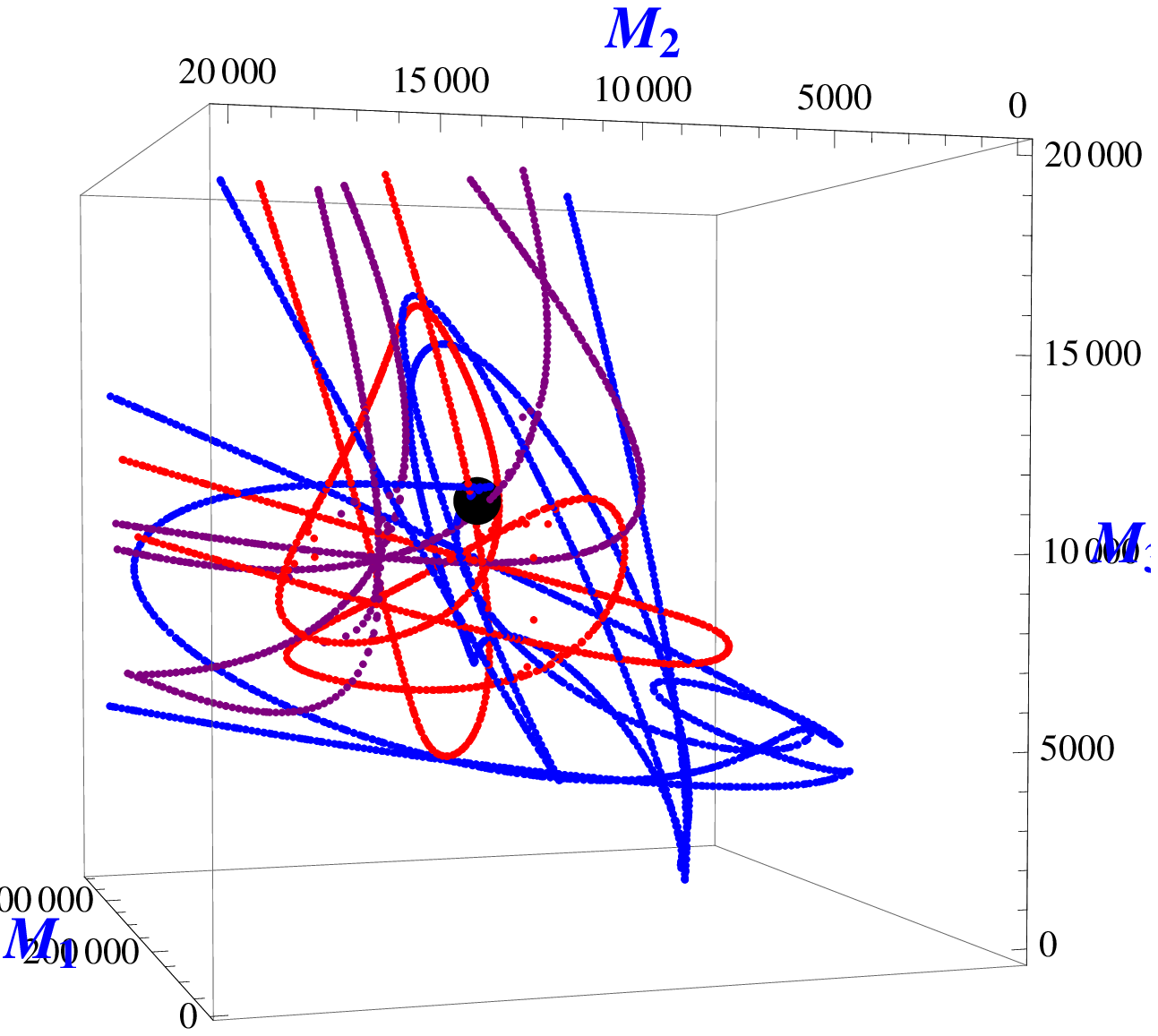}
\includegraphics[width=50mm]{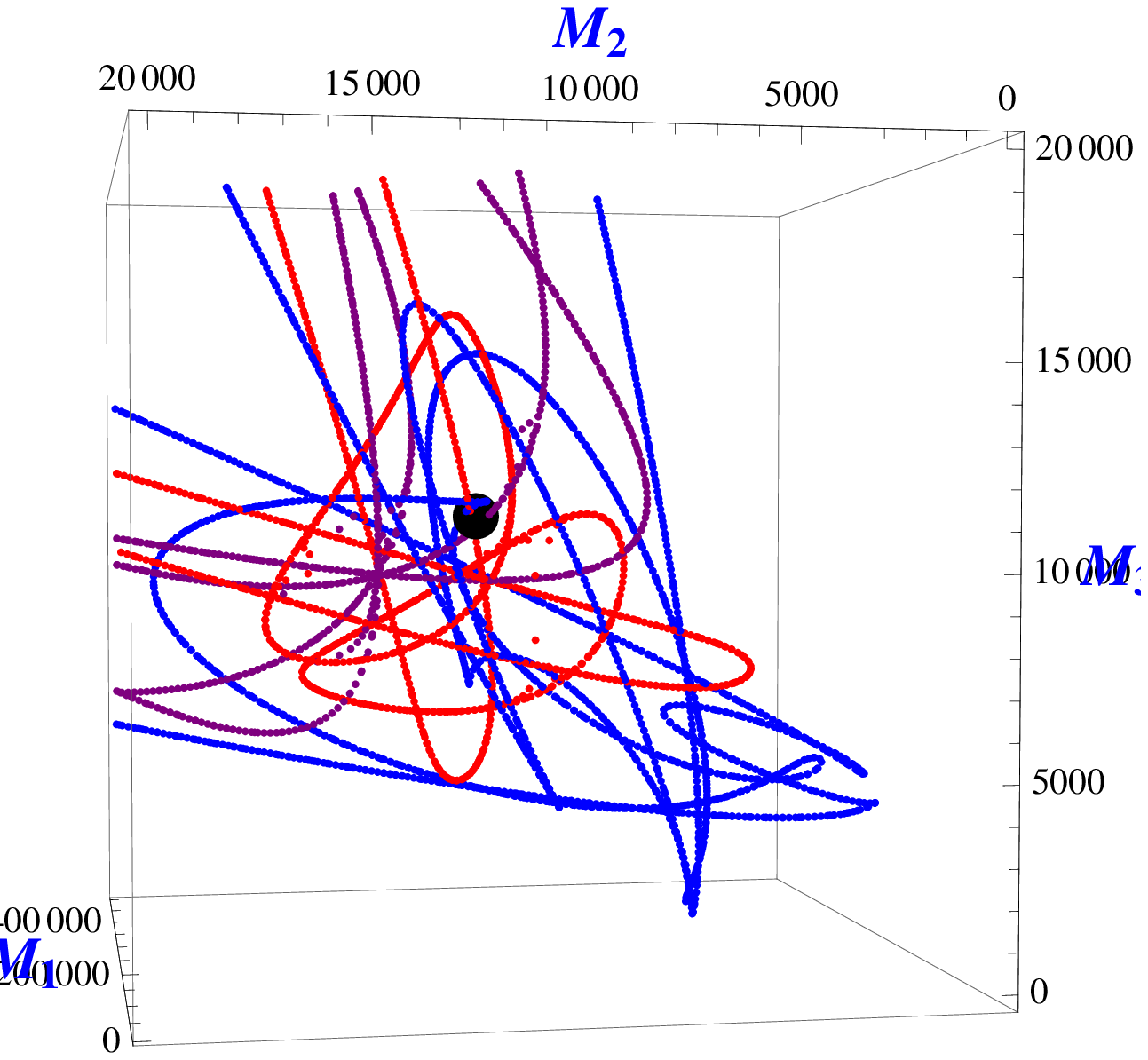}
\caption{Stereoscopic views of the parton-level solution curves for
  the same three events, now including all
  combinations.}\label{fig:mass_tru3Dp}
\end{center}\end{figure}

Figure~\ref{fig:mass_tru3Dp} shows the effect of combinatorial
ambiguities for the same three events, viewed from a different angle
for clarity.  Here the interchanges of near and far leptons
($2\leftrightarrow 3$ and $6\leftrightarrow 7$) and of quarks
($1\leftrightarrow 5$) are included, making eight combinations per
event.  Three-dimensional viewing reveals that incorrect combinations
either have no real solutions or tend to give curves that do not
congregate to form regions of high density.

In the real world, the effects of parton showering, hadronization and detector
resolution shift and distort the solution curves. The density of
solutions around the TMP is reduced, and incorrect combinations may
happen to produce other regions of high density.  In addition, QCD
radiation produces extra jets, which increase the number of wrong
combinations.

To investigate these effects, we study an inclusive SUSY sample containing
SUSY backgrounds as well as signal processes, generated 
using {\small HERWIG} 6.5 with initial and final state radiation turned on.
The sample is interfaced with AcerDET 1.0~\cite{RichterWas:2002ch}.  Final-state  
hadrons are formed into jets, and the momenta of jets and leptons
are smeared according to the simulated detector resolution.

\begin{table}[!t]
  \centering
  \begin{tabular}{|c||c|c|c||c|c|c|c|}
    \hline & $m_0$ & $m_{1/2}$ & $A_0$ &
    $\tilde{\chi}_1^0$ & $\tilde{e}_R$ & $\tilde{\chi}_2^0$ & $\tilde{u}_L$ \\ \hline
    \hline Point A & 110 & 220 & 0 & 86 & 142 & 161 & 504 \\ \hline
    Point B & 100 & 250 & $-100$ & 99 & 141 & 186 & 563 \\ \hline
    Point C & 140 & 260 & 0 & 103 & 174 & 193 & 592 \\ \hline
  \end{tabular}
  \caption{Parameters and mass spectra in GeV for non-CMSSM model points A, B and C. 
   Parameters common to all points are  
  $m_0^{\rm 3rd\,gen.}=300$ GeV, $\tan\beta=10$, ${\rm sign}(\mu)=+$.}
  \label{tab:spectra}
\end{table}

To obtain a larger event sample with two decay chains like (\ref{eq:chain_susy}),
for this analysis we adopt non-CMSSM model points A, B and C, where the third-generation soft mass is 
larger than the others, so that the branching ratio (\ref{eq:chain_susy}) 
is increased by suppressing the $\tilde \chi_2^0 \to \tilde \tau_1^{\pm} \tau^{\mp}$ mode.
The sparticle spectra at these points are shown in
Table~\ref{tab:spectra}.  The generated samples of 500,000 events
correspond to about 10, 15 and 20 ${\rm fb}^{-1}$ of integrated luminosity, respectively.

The following cuts are applied in order to select signal events:
\begin{description}
\item[ (i)] $M_{\rm eff} \equiv \sum_{i=1}^{4} p_T^{{\rm jet},i}  
 + \sum_{i=1}^4 p_T^{{\rm lep},i} + E_T^{\rm miss}> 400\, {\rm GeV}$ \,;
\item[ (ii)] $E_T^{\rm miss} > \max( 200\,{\rm GeV},~0.2M_{\rm eff})$\,;
\item[ (iii)] At least two jets with $p_T^{{\rm jet},1} > 100\,{\rm GeV}$ and 
$p_T^{{\rm jet},2} > 50\,{\rm GeV}$ within $| \eta |< 2.5$\,;
\item[ (iv)] Two pairs of opposite sign same flavour leptons with $p_T>20\,{\rm GeV}$ and $|\eta|<3$\,;
\item[ (v)] No $b$ jet with $p_T>30\,{\rm GeV}$ and $|\eta|<3$\,.
\end{description}
The $b$ tagging efficiency is assumed to be $60\%$.
In the cut {\bf (iv)}, we select not only opposite-flavour lepton pairs ($e^+e^- \mu^+ \mu^-$) 
but also the same-flavour pairs ($e^+e^-e^+e^-$ and $\mu^+\mu^-\mu^+\mu^-$) 
to have larger samples, although the latter have double the combinatorial background of the former.
If an event contains more than two hard jets, we take the three hardest jets as candidates
for the jets from the signal decay chains (\ref{eq:chain_susy}), and try all possible combinations. 
The number of combinations is 8 (16) for two candidate jets and 24 (48) for three 
with opposite (same) flavour lepton pairs.
The numbers of events that survive the above cuts are shown in the first row in Table \ref{tab:results}
together with signal/background ratios for each model point.
The background is rather mixed, coming mainly from direct $\tilde
\chi_2^0$ productions associated with squarks or gauginos as well as
modes containing $\tilde q_R \to \tilde \chi_2^0 j$, $\tilde b_1 \to
\tilde \chi_2^0 b$ and $\tilde \chi_2^0 \to \tilde \chi_1^0 l^+ l^-$.
For model point C, the three-body decay $\tilde \chi_2^0 \to \tilde
\chi_1^0 l^+ l^-$ is enhanced  because $m_{\tilde \chi_2^0} \simeq m_{\tilde \chi_1^0} + m_Z$
and turns out to be the main background.
Standard Model background is expected to be negligible after the above selection
cuts.  According to ref.~\cite{Ball:2007zza}, the potential background comes from 
$t \bar t \to b \bar bW^+ W^- \to 4l$.  Based on {\small HERWIG} 6.5
simulation of this process, we confirmed that it is indeed negligible after cuts.

\begin{table}[!t]
  \centering
  \begin{tabular}{|c||c|c|c|}
    \hline & Point A & Point B & Point C  \\ \hline \hline
    Events (S/B) 
    & 326 (4.2) & 499 (4.5) & 292 (2.8)  \\ \hline
    Sharing (S/B)
    & 219 (8.1)  & 341 (9.7) & 172 (4.9) \\ \hline
$M_1$ (True ; Best) & 231890 ; $222500$ & 286157 ; $282500$ & 316274 ; $317500$ \\ \hline 
$M_2$ (True ; Best) & 5624 ; $5000$ & 14520 ; $14200$ & 6815 ; $6600$ \\ \hline
$M_3$ (True ; Best) & 12872 ; $11700$ & 10293 ; $9900$ & 19812 ; $18900$  \\ \hline
  \end{tabular}
  \caption{
  First row:  number of events (signal/background) after cuts.  
  Second row:  number of events that contribute to the best-fit cell in the $\Delta \chi^2$
  distribution.
  Third to fifth rows:  true mass and the central value of the best-fit cell in {$\rm GeV^2$}.
  }
  \label{tab:results}
\end{table}

If the detector and jet properties are well understood, 
from the observed jet momentum, $p^{\rm jet}$, 
we may stochastically estimate the original parton momentum,
$p^{\rm par}$, with a gaussian distribution $\epsilon (p^{\rm par} | p^{\rm jet})$.
In this situation, we can built a confidence region in the ($M_1$, $M_2$, $M_3$) space \cite{Kawagoe:2004rz}.
For each signal event combination, $i_{\rm ev}$,
a probability density function may be constructed as
\beqn
f_{i_{\rm ev}} ({\bf M}) = \frac{1}{N_{i_{\rm ev}}} 
\int dp^{\rm par}_1  dp^{\rm par}_2 \epsilon (p^{\rm par}_1 | p^{\rm jet}_1) \epsilon (p^{\rm par}_2 | p^{\rm jet}_2)
\delta(p_4^2-m^2_N) \delta(p_8^2-m^2_N),
\eeqn
where $p_4$, $p_8$ and $m_N$ are the functions of ${\bf M}$ and $p^{\rm par}_{1,2}$
given in section \ref{sec:method}, and $N_{i_{\rm ev}}$ is a normalization factor. 
Given $N$ event-combinations, log-likelihood and $\Delta \chi^2$ functions are obtained as
\beq
\ln L({\bf M}) = \sum_{i_{\rm ev}}^N  \ln f_{i_{\rm ev}} ({\bf M})
\eeq
and
\beq
\Delta \chi^2 ({\bf M}) = 2( \ln L({\bf M})_{\rm max} -\ln L({\bf M})),
\eeq
respectively, 
where $\ln L({\bf M})_{\rm max}$ is the maximum value of $\ln L({\bf M})$ in the space ${\bf M}$.

We calculate $\ln L({\bf M})$  approximately by the following procedure.
For each event, we generate Monte Carlo ``fake" events whose jet momenta are shifted 
from the original ones according to the probability distribution $\epsilon(p^{\rm par}|p^{\rm jet})$.
The parameter space ${\bf M}$ is divided into cells.
For each cell, we count the number of fake events for which the
solution curves go through that cell. 
If different combinations of the same event yield two or more curves passing through the same cell,
we count only one.
If the number of fake events is large and the cell size is small,
this provides $f_{i_{\rm ev}}({\bf M}_{\rm cell})$ with a certain normalization.
As long as we work with $\ln L({\bf M})$, the normalization factor $N_{i_{\rm ev}}$ is irrelevant, 
because it only shifts the constant term of $\ln L({\bf M})$. 
We ignore cells that have $f_{i_{\rm ev}}({\bf M}_{\rm cell}) = 0$ in
our log-likelihood calculation,
setting $\ln f_{i_{\rm ev}}({\bf M}_{\rm cell}) = 0$.
Finally, we sum up $\ln f_{i_{\rm ev}}({\bf M}_{\rm cell})$ for all
combinations of all events.

\begin{figure}[t!]
\begin{tabular}{lr}
\hspace{-13mm}
\begin{minipage}{0.3\hsize}
\begin{center}
(A)
\includegraphics[width=55mm]{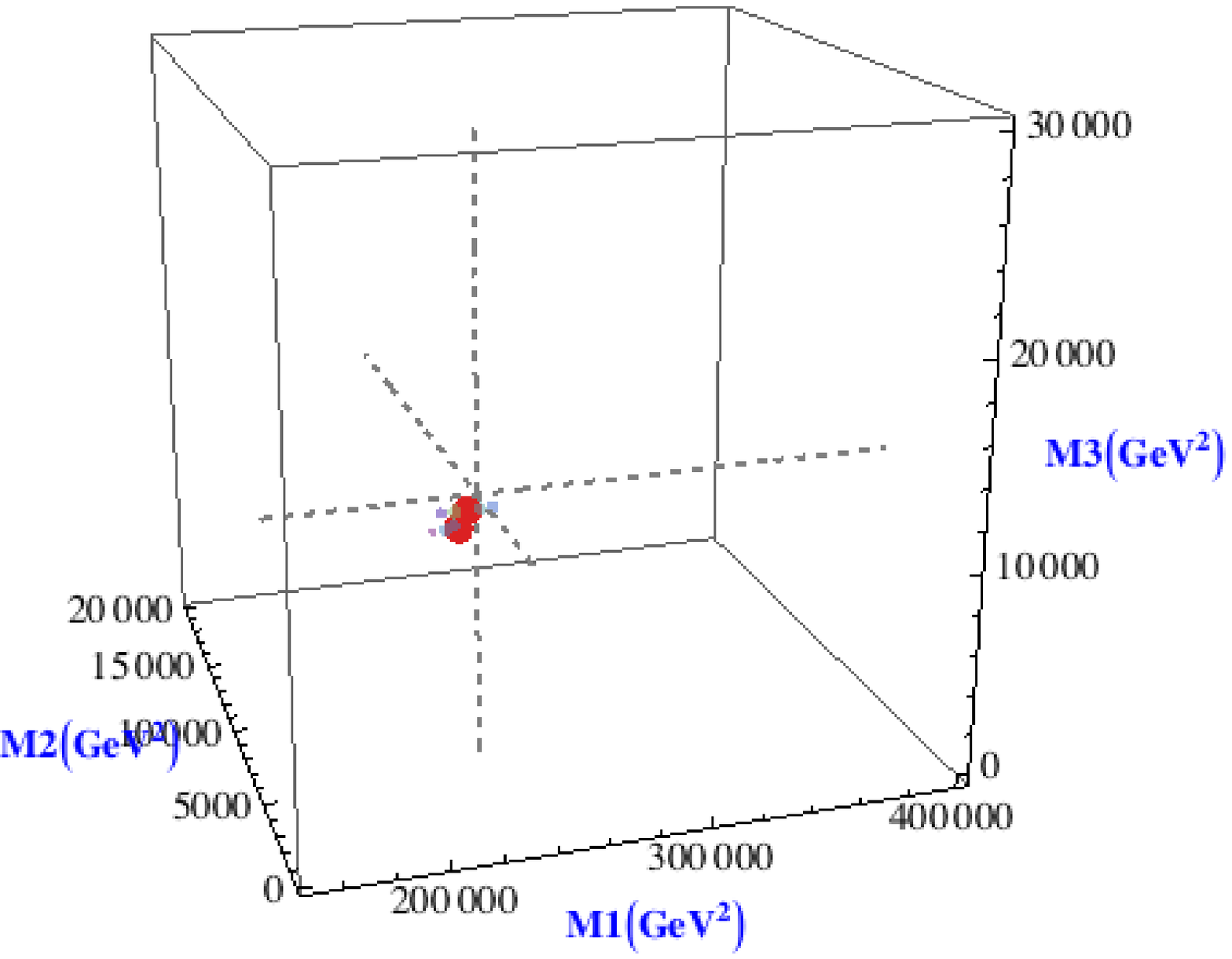}
\end{center}
\end{minipage}
\hspace{3mm}
\begin{minipage}{0.3\hsize}
\begin{center}
(B)
\includegraphics[width=55mm]{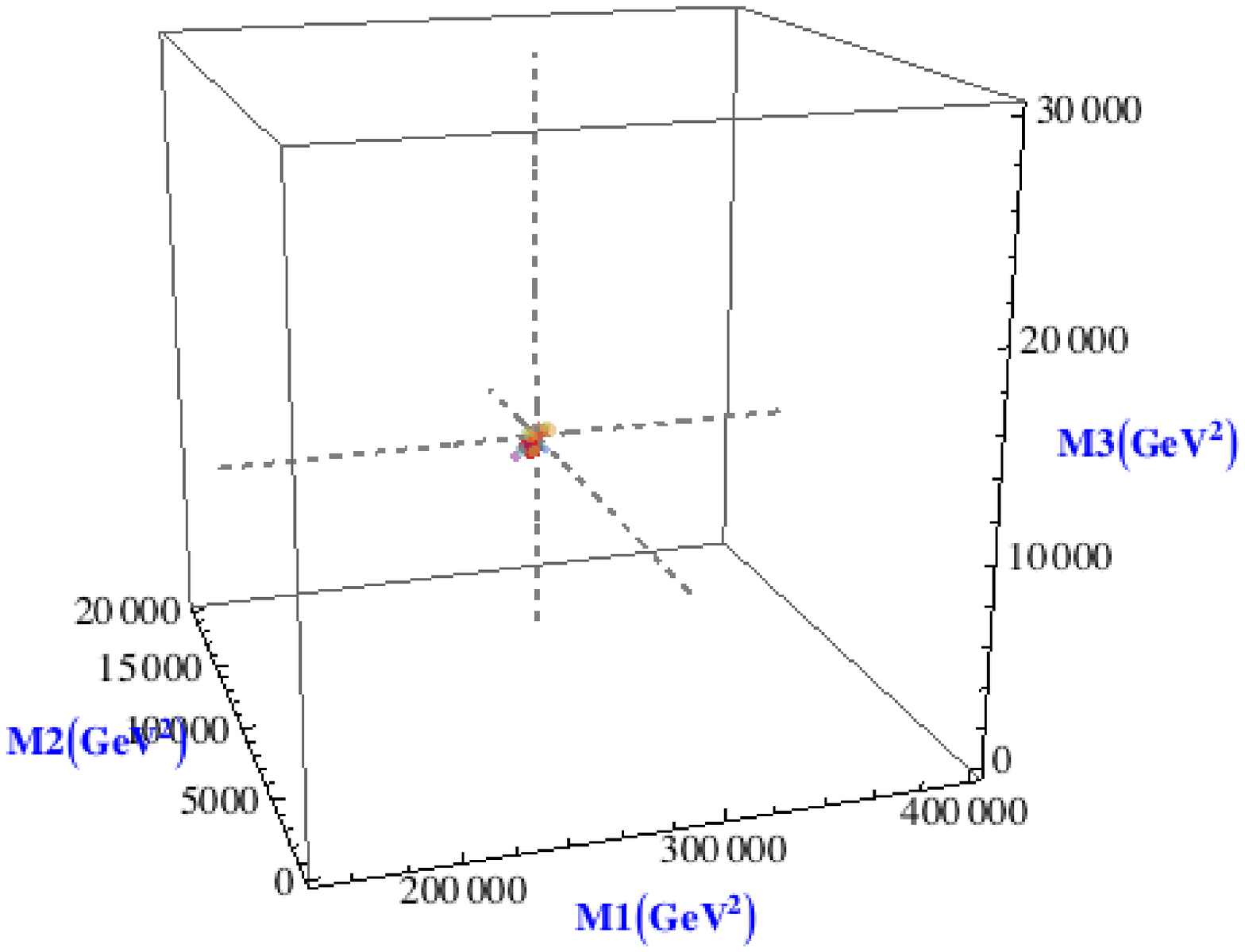}
\end{center}
\end{minipage}
\hspace{4mm}
\begin{minipage}{0.3\hsize}
\begin{center}
(C)
\includegraphics[width=55mm]{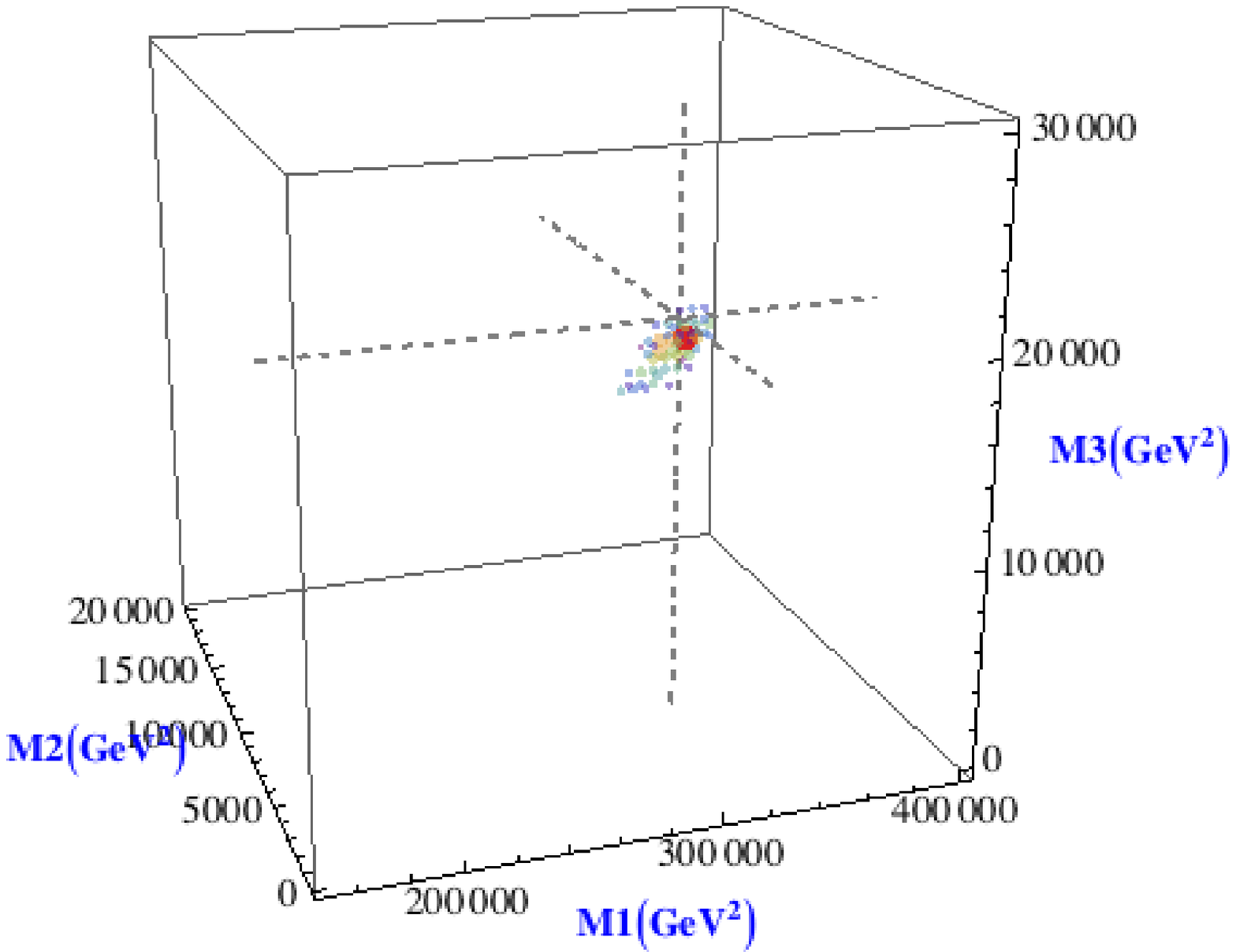}
\end{center}
\end{minipage}
\hspace{-11mm}
\begin{minipage}{0.3\hsize}
\begin{center}
\includegraphics[width=10mm]{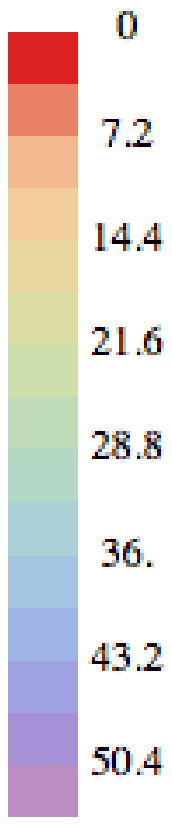}
\end{center}
\end{minipage}
\end{tabular}
\caption{
 Distribution of $\Delta \chi^2({\bf M})$ for each model point at detector level.
 The true mass point is at the intersection of the three dashed lines.
}
\label{fig:likely}
\end{figure}

In the following analysis,
we generate 1000 Monte Carlo fake events for each event.  
For the smearing of jets and the missing transverse momentum, we use gaussian functions 
with the following standard deviations,
obtained by parametrizing the AcerDET results:
\beqn
\frac{\sigma_E}{E} = \frac{0.5}{\sqrt{E}} + 0.03,  ~~~~~
\sigma_{\phi} = \frac{0.4}{\sqrt{E}} + 0.015,  ~~~~~
\sigma_{\eta} = \frac{0.3}{\sqrt{E}} + 0.02,
\eeqn
for jets and 
\beqn
\frac{\sigma_E}{E} = \frac{0.5}{\sqrt{E}} + 0.03,  ~~~~~~
\sigma_{\phi} = \frac{0.8}{\sqrt{E}} + 0.06,  
\eeqn
for the missing transverse momentum.
We do not smear the lepton momenta because
mismeasurement of lepton momenta is negligible
compared to the jet smearing.

Figure \ref{fig:likely} shows the $\Delta \chi^2 ({\bf M})$ distribution obtained by the above procedure
for each model point.
The cell size is $\Delta M_1 = 5000$, $\Delta M_2 = 400$, $\Delta M_3 = 600$ in ${\rm GeV}^2$. 
The distribution has only one sharp minimum, 
which is close to the TMP, as can be seen in Table~\ref{tab:results}.
Backgrounds from wrong combinations and different decay chains 
do not produce local minima at other places, 
and the effect of those backgrounds may be less significant around the true mass point.  

The second row in Table~\ref{tab:results} shows how many different events share the best-fit cell;
the signal/background ratios in that cell are also shown in parentheses.  
The ratios are improved significantly.
For each model point the ratio is about twice that for the whole sample.
  
\begin{table}[!t]
  \centering
  \begin{tabular}{|c||c|c|c|c|}
    \hline & $\tilde{\chi}_1^0$ & $\tilde{e}_R$ & $\tilde{\chi}_2^0$ & $\tilde{u}_L$ \\ \hline \hline
    Point A &  $68.2^{+16.2}_{-5.8}$ & $127.9^{+12.6}_{-4.2}$ & $146.1^{+13.0}_{-4.4}$ & $493.8^{+11.5}_{-3.8}$ \\ \hline
    Point B & $94.5^{+8.5}_{-2.8}$ & $137.2^{+9.1}_{-3.1}$ & $181.7^{+8.5}_{-2.8}$ & $561.7^{+9.4}_{-3.1}$ \\ \hline
    Point C & $95.6^{+5.1}_{-5.3}$ & $167.4^{+3.9}_{-3.9}$ & $186.1^{+4.0}_{-4.0}$ & $593.4^{+3.4}_{-3.4}$ \\ \hline
  \end{tabular}
  \caption{Estimated sparticle masses with their errors in GeV. 
  }
  \label{tab:mass_results}
\end{table}

In the third to fifth row of Table~\ref{tab:results}, we show the central values of the best-fit cells
compared to the TMP at each model point.
As can be seen, the best-fit points are slightly biased towards lower masses.
This may result from the following systematic errors in the present analysis.
First, the AcerDET jets that we use are defined as massless, whereas
the 4-momenta defined by $p^{\rm par}=p(\tilde q) - p(\tilde \chi^0_2)$ have masses 
of around 10-100 GeV after fragmentation and hadronization.
Second, we have parametrized the probability distributions of parton
momenta by gaussian functions.
However, the difference between a parton momentum in the event record and the AcerDET jet momentum
deviates slightly from a gaussian distribution, due to the underlying event, hadronization effects and 
high-$p_T$ gluon emission from the original parton.
A better jet algorithm with jet masses and a more refined
parametrization will be needed to reduce these systematic errors.

Table~\ref{tab:mass_results} shows the sparticle masses estimated from our analysis.
The errors are obtained from $1\sigma$ regions assuming the errors in
$M_1$, $M_2$ and $M_3$ are uncorrelated,
where the $1\sigma$ region is defined by $\Delta \chi^2 < 3.53$.
We neglect the error from the mismeasurement of the dilepton endpoint
because of its expected good accuracy.
The $1\sigma$ errors at Point C are accidentally small despite the
small number of signal events compared to the other points.  The
sizable errors also come from the cell size,
because the $1\sigma$ region is almost the same size as the cells.
More precise error estimation would require smaller cells.
In addition, larger numbers of fake Monte Carlo events could be used
to make the probability density smooth around the peak
region.  Such refinements would be justified and straightforward in
an analysis of real data.

\begin{figure}[t!]
\begin{tabular}{lr}
\hspace{-8mm}
\begin{minipage}{0.33\hsize}
\begin{center}
(A)
\includegraphics[width=58mm]{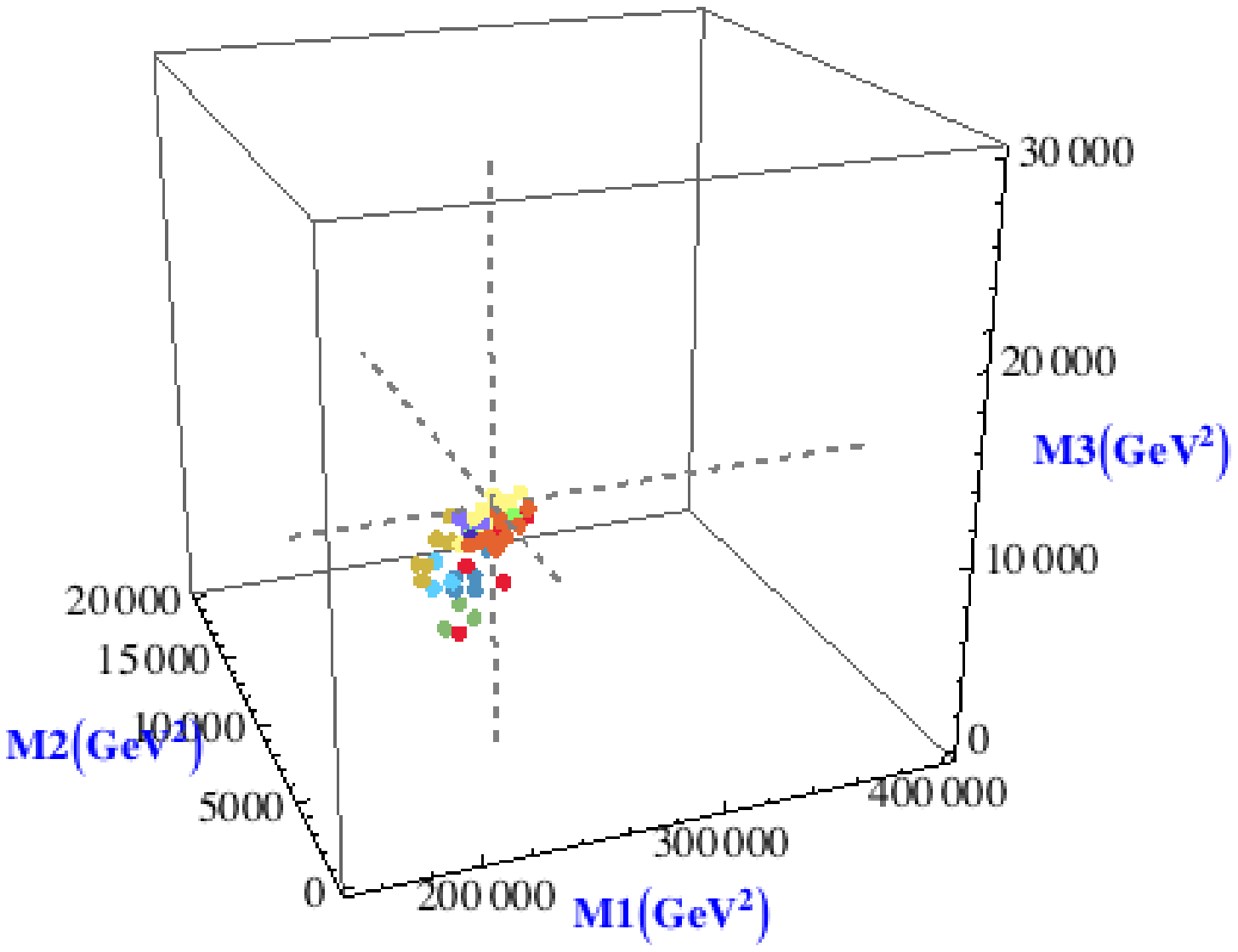}
\end{center}
\end{minipage}
\hspace{1mm}
\begin{minipage}{0.33\hsize}
\begin{center}
(B)
\includegraphics[width=55mm]{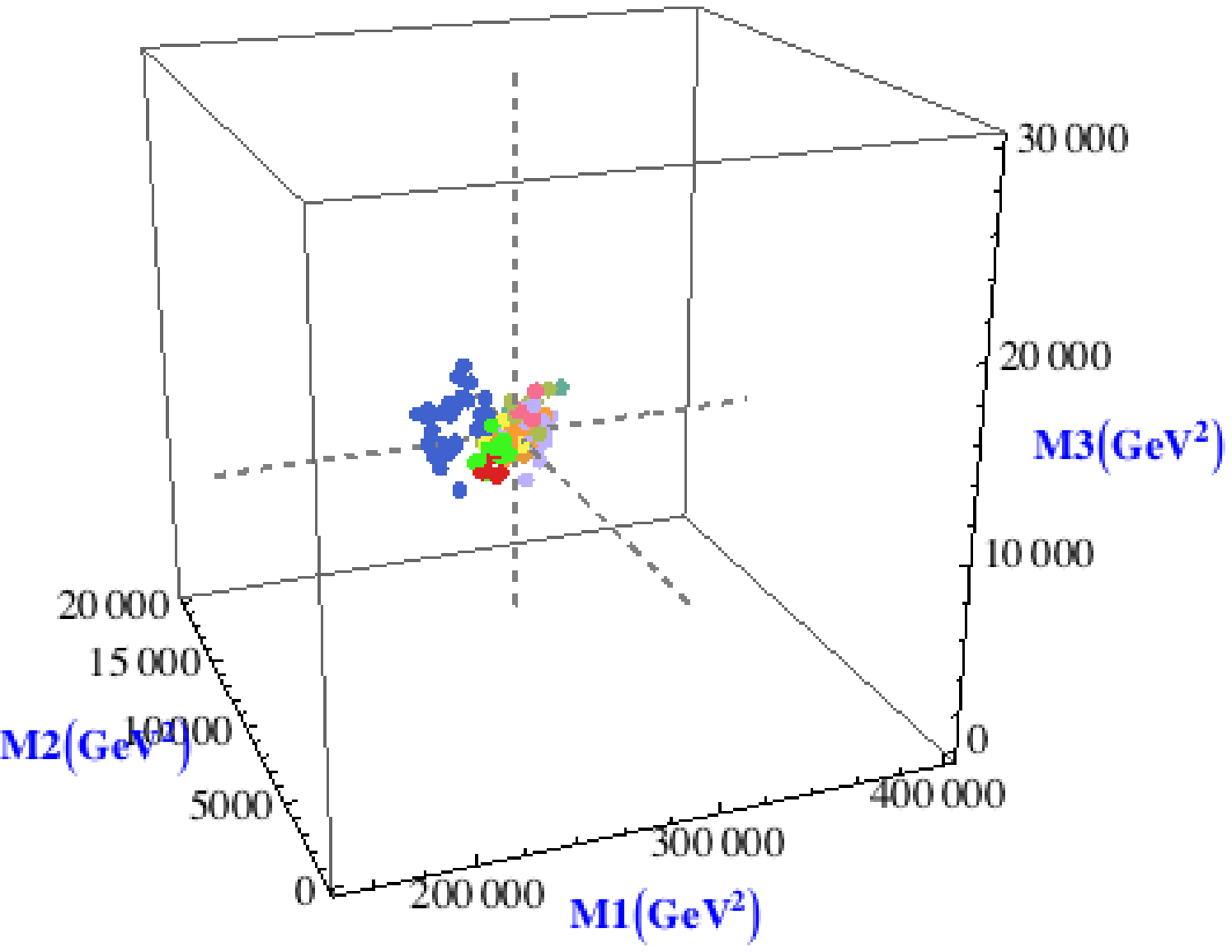}
\end{center}
\end{minipage}
\hspace{2mm}
\begin{minipage}{0.33\hsize}
\begin{center}
(C)
\includegraphics[width=55mm]{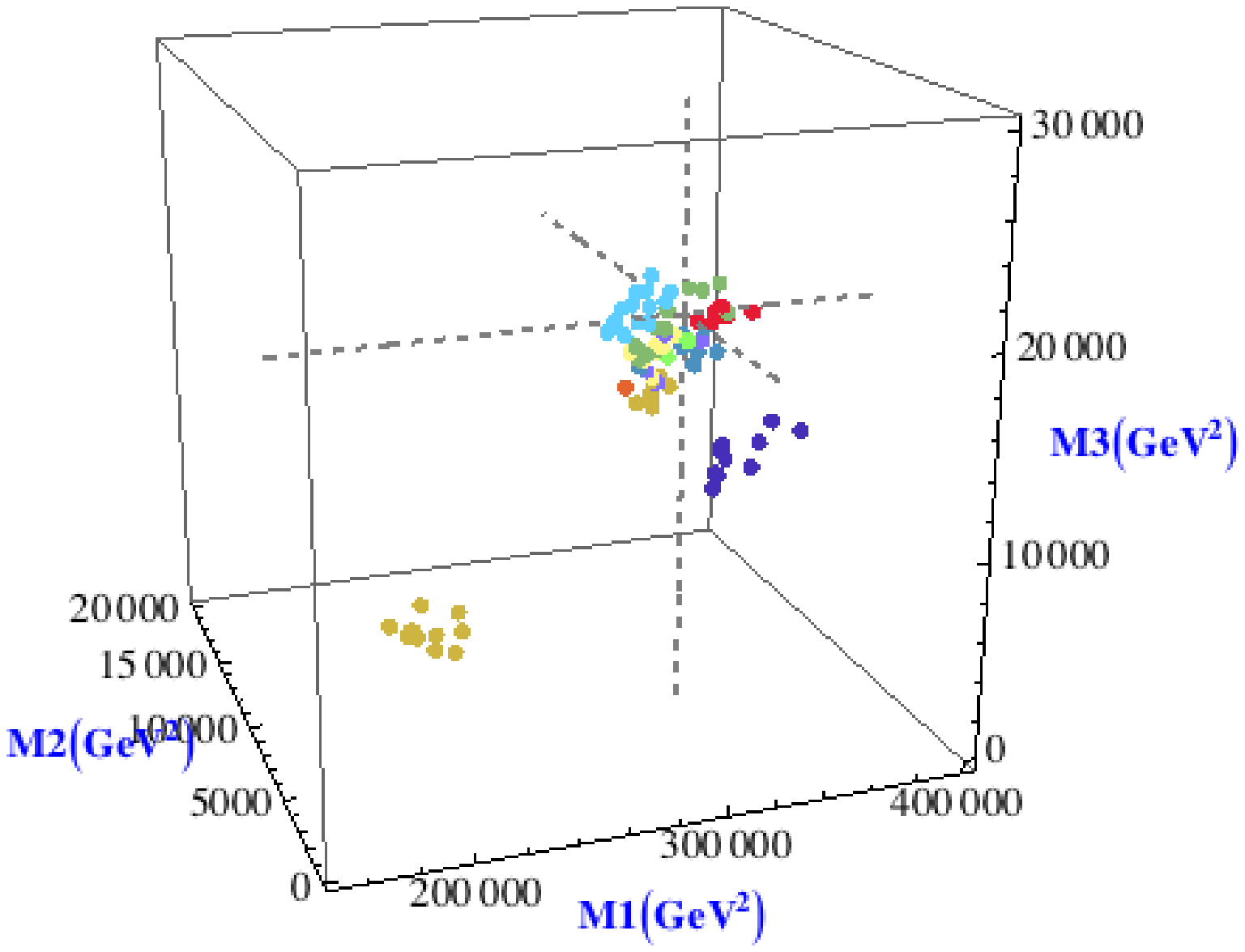}
\end{center}
\end{minipage}
\end{tabular}
\caption{ One-sigma regions of 10 sub-samples, distinguished by their colours. 
Each sub-sample contains 25 events.  
}
\label{fig:subsets}
\end{figure}

The statistical approach we have adopted here is justifiable for large event samples.
In order to see what happens in small samples, 
and to check the interpretation of $\Delta \chi^2 $,
we divide the samples after cuts 
into several sub-samples, so that each of them contains 25 events.
Figure \ref{fig:subsets} shows the $1 \sigma$ regions of 10 sub-samples for each model point.
The sub-samples are distinguished by their colours.  As expected,
the $1\sigma$ regions are more widely spread compared to Figure 
\ref{fig:likely}.   
As can be seen in Figure \ref{fig:subsets}\,C, 
the gold-coloured sub-sample has  two local minima, one of them away from the TMP.
Furthermore, the $1\sigma$ region of the blue-coloured sub-sample is localised away from the TMP. 
Those sub-samples have fewer signal events compared to the others.
The signal/background ratio is 1.8 (2.6) for the blue (gold) sub-sample.  
We checked, however, that the $2 \sigma$ region of the blue sub-sample contains the TMP.  
Despite the small sub-sample sizes, the maximum-likelihood regions are still mainly
localised around the TMP,
and their average sizes scale as expected with the number of events.
In our approach, unlike in endpoint and kink methods, all signal events contribute
to the determination of the unknown masses,
no matter where they may lie in phase space, and so the method can
provide meaningful information about the unknown masses even in rather small samples.

\section{Conclusions}

The method of mass determination presented above is simple to apply and looks
promising for the class of processes studied here. 
We demonstrated the validity of the method by means of full simulations including
detector effects.  
Combinatorial background, the background from other SUSY processes and
the effects of additional jets due to QCD radiation 
do not appear to be a serious problem.
A statistical approach is applicable to deal with
jet momentum mismeasurement. 
We constructed an effective $\Delta \chi^2$ variable 
which allows a rather precise determination of the unknown masses with
controlled statistical errors.  There are identified systematic
errors, leading to a bias towards lower masses,  which could be
reduced with an appropriate jet algorithm and improved parametrization
of jet momentum smearing. The method can be applied successfully even to small event samples,
because it makes full use of the kinematical information from every event. 

{\it Note added:} The procedure adopted here for constructing an
approximate likelihood function by generating large numbers of
``fake'' events is quite time-consuming.  An analytical procedure
similar to that outlined (but not implemented) for the exact-solution
method in Section 6 of ref.~\cite{Cheng:2009fw} may be applicable and
more efficient.  We thank H.~C.~Cheng for this suggestion.

\section*{Aknowledgements}
KS and BW are grateful for helpful discussions with other members of
the Cambridge SUSY Working Group.  BW thanks the CERN Theory
Group for hospitality during part of this work.


\end{document}